\input harvmac
\input epsf

\def\p{\partial}

\def\half{{1\over 2}}

\def\a{\alpha}
% \def\lt{\tilde{\lambda}}

%special

\def\c{\theta}

%%%%%%%%%%%%%%%%%%%%%%%%%%%%%%%%%%%%%%%%%%%%%%%%%%%%%%%%%%%%%%%%%%%%%

\Title{}{\vbox{\centerline{Shock Waves and Cosmological Matrix Models}}}

\centerline{Miao Li$^{1,2}$ and Wei Song$^{1}$}

\medskip

\centerline{\it $^1$ Institute of Theoretical Physics}
\centerline{\it Academia Sinica, P. O. Box 2735} \centerline{\it
Beijing 100080}
\medskip
\centerline{\it and}
\medskip
\centerline{\it $^2$ Interdisciplinary Center of Theoretical
Studies} \centerline{\it Academia Sinica, Beijing 100080, China}
\medskip

\centerline{\tt mli@itp.ac.cn} \centerline{\tt wsong@itp.ac.cn}
\medskip

We find the shock wave solutions in a class of cosmological
backgrounds with a null singularity, each of these backgrounds
admits a matrix description. A shock wave solution breaks all
supersymmetry meanwhile indicates that the interaction between two static
D0-branes cancel, thus provides  basic evidence for the matrix
description. The probe action of a D0-brane in the background of
another suggests that the usual perturbative expansion of matrix
model breaks down.

\Date{July 2005}

%\draft

\nref\csv{B. Craps, S. Sethi and E. Verlinde,``A Matrix Big Bang,"
hep-th/0506180.}
\nref\mli{M. Li, ``A class of cosmological matrix models," hep-th/0506260.}
\nref\hs{G. T. Horowitz and A. Steif, ``Singular string solutions with non-singular
initial data," Phys. Lett. B258 (1991) 91.}
\nref\lms{H. Liu, G. Moore and N. Seiberg, ``Strings in a Time-Dependent Orbifold,"
hep-th/0204168, JHEP 0206 (2002) 045; hep-th/0206182, JHEP 0210 (2002) 031.}
\nref\alor{ A. Lawrence, ``On the instability of 3d null singularities,"
hep-th/0205288, JHEP 0211 (2002) 019.}
\nref\hp{ G. T. Horowitz and J. Polchinski, ``Instability of Spacelike and Null Orbifold
Singularities," hep-th/0206228, Phys.Rev. D66 (2002) 103512.}
\nref\bckr{ M. Berkooz, B. Craps, D. Kutasov and G. Rajesh, ``Comments on Cosmological
Singularities in String Theory," hep-th/0212215, JHEP 0303 (2003) 031.}
\nref\bfss{T. Banks, W. Fischler, S. Shenker and L. Susskind, ``M theory as A Matrix
Model: A Conjecture," hep-th/9610043, Phys. Rev. D55 (1997) 5112.}
\nref\ls{L. Susskind, ``Another Conjecture about Matrix Theory," hep-th/9704080.}
\nref\becker{K.~Becker, M.~Becker, J. Polchinski and A. Tseytlin,
``Higher Order Graviton Scattering in M(atrix) Theory," Phys. Rev. D56 (1997) 3174,
 hep-th/9706072.}

It is an important open problem to formulate string/M theory in a time-dependent
background in general, and to study what happens to cosmological singularity in
particular. The recently proposed matrix description \csv\ of a cosmological background
with a null singularity appears to be an advance forward, this construction was generalized
to a class of backgrounds in \mli.
An earlier example of cosmology with a null singularity
is proposed in \refs{\hs,\lms} and is later studied by many authors \refs{\alor-\bckr},
they find that this singularity is highly unstable due to gravitational back-reaction.
It seems to us that models studied in \refs{\csv,\mli} and in the present note are devoid
of such problems.

A cosmological background with a null singularity is not necessarily stable, even when it
admits a perturbative string description \refs{\alor-\bckr}. It was found that a smeared
shock wave changes the nature of the singularity, this happens in $2+1$ dimensions. Our
solution presented below is also smeared in one or more directions, but localized in
other directions, there is no sign of instability.

Matrix theory \refs{\bfss, \ls} description predicts that there is no interaction between
two static D0-branes,
this is verified by our solution. The matrix action does not possesses linear supersymmetry
(there is always non-linearly realized supersymmetry corresponding to shift of fermions),
we show that the shock wave solution breaks all supersymmetry, in accordance with the
matrix action prediction. Does the fact that there is no net force between two static
D0-branes imply that we can expand the effective action in terms of relative velocity?
The probe action in the background of a shock wave shows that this is not the case. This
has far-reaching consequence: the usual perturbative expansion in matrix model breaks
down.

The background studied in \csv\ is a flat 10D space-time with a linear dilaton proportional
to the light-cone coordinate $x^+$. Its 11 dimensional M theory matric reads
\eqn\mtmet{ds_{11}^2=e^{ 2 Q x^+/3 } ds_{10}^2 + e^{-4 Q x^+/3} (dY)^2.}
The 10 dimensional part shrinks when $x^+\rightarrow -\infty$, thus there is a null
singularity a finite distance away in terms of the affine parameter. It is argued in \csv\
that the matrix model works for this background with a time dependent action, where
the matrix time $t$ is chosen to coincide with $x^+$.

The above solution can be generalized to the following class \mli
\eqn\mmtm{\eqalign{ ds^2&=e^{2\a x^+}(-2dx^+dx^-+(dx^i)^2)+e^{2\beta x^+}(dx^a)^2\cr
i&=1,\dots, 9-d,\quad a=9-d+1,\dots,9,}}
where the transverse directions are grouped into two sets.
There are other generalizations in \mli, but here we restrict our discussion to the above two cases.

It is believed that for each of these backgrounds there is a matrix description, which may provide a theory
for resolving the singularity. For this claim to be valid, it is important to check whether
these backgrounds are stable.
Here we take an approach similar to that of \alor.
Consider there be some gravitational sources,
localized in $x^-$ dimension as well as in the transverse dimensions.
The existence of such sources will change the background.
We want to see how large the change can be.
First we consider the easiest case, with the background metric given by \mtmet:
It is convenient to treat the $x^-$ and $Y$ dimension as periodically identified,i.e.
$$x^-\sim x^-+2\pi R,\quad Y\sim Y+2\pi R'$$

Consider the perturbation represented by stress tensor
\eqn\pstr{
    T_{--} =\sum_{a}p_{-a} \  e^{-2Q x^+}\delta({\vec{x}}_{\bot}-{\vec{x}}_{\bot a}),}
where the $x^{+}$ dependence is introduced by energy conservation (this is for the
background \mtmet; for \mmtm, the $x^+$ dependent factor is
$\exp[-((9-d)\a+d\beta)x^+]$), and
$x_{\bot}$ represents all the uncompactified  transverse dimensions.
The above stress tensor is due to multi-sources,
located at ${x_{\bot}}_{a}$ in the transverse directions. The stress tensor is smeared in
the $x^-$ and $Y$ directions, thus $p_{-a}$ is not really the longitudinal
momentum of the $a$-th particle, the longitudinal momentum should be $p_{-a}2\pi R
2\pi R'$, where $R$ is the radius of $x^-$ and $R'$ the radius of $Y$.
We make the ansatz that the metric takes the following form
\eqn\anssl{
   ds^2=-2e^{2\alpha x^{+}}dx^{+}dx^{-}-2e^{\theta
   x^{+}}f(x^-,x^{A})(dx^{-})^2+e^{2\alpha_{A}x^{+}}(dx^{A})^2,}
where $A=i, a$ in \mmtm, $\theta$ is a constant to be determined. This ansatz can solve
Einstein equations with source \pstr,
and requires more non-vanishing components in stress tensor if the source is not
smeared in direction $Y$ (or in directions $x^a$), as we shall see shortly.
The component $g_{--}$ reflects the effect the source has on the geometry.
If $g_{--}$ vanishes, we get back to the original metric we started with, which
should satisfy the vacuum Einstein equations with the constraint on the parameters:
\eqn\eomf{
\sum_A\a_A(\a_A-2\a)=0.}
For metric \mtmet, we have
\eqn\mone{\a=\a_i=Q/3,\quad i=1\ldots8;\quad\a_{9}=-2Q/3.}
In the case of \mmtm, there are
\eqn\mtwo{
\a_i=\a,\quad\a_{a}=(1\pm3/\sqrt d)\a.}
To find a solution with a source \pstr, we use the orthonormal basis
\eqn\tetr{\eqalign{
  e^{-}& = dx^{-} \cr
  e^{+} &={e^{2\alpha x^{+}}dx^{+}+e^{\theta x^{+}}dx^{-}}  \cr
  e^A &=e^{\alpha_Ax^{+}}dx^{i}.}}
Up to symmetry, the non-vanishing components of spin connection are
\eqn\conn{\eqalign{
  {\omega^{+}} _{+}&=\theta e^{(\theta-2\alpha)x^{+}}dx^{-}\cr
  {\omega^{+}} _{A}&=e^{(\theta-\alpha_{A})x^{+}}\partial_A f dx^{-}
  -\alpha_{A}  e^{(\alpha_{A}+\theta-2\alpha)x^{+}}f dx^{+}\cr
   {\omega^{A}}_{+}&=\alpha_{A} e^{(\alpha_{A}-2\alpha)x^{+}}dx^{A}.}}
The non-vanishing curvature 2-forms are
\eqn\curf{\eqalign{
{\Omega^{+}}_{+}&=\theta(\theta-2\alpha)e^{(\theta-4\alpha)x^{+}}fe^{+}\wedge e^{-} +
(\theta-\alpha_{A})  e^{(\theta-2\alpha-\alpha_{A})x^{+}}\partial_{A}f e^{i}\wedge e^{-},\cr
{\Omega^{+}} _{A}&=[e^{(\theta-\alpha_{A}-\alpha_{A})x^{+}}\partial_{A}\partial_{B}f
-\alpha_{A}(\alpha_{A}-2\alpha)e^{(2\theta-4\alpha)x^{+}}f^2\delta_{AB}
+\alpha_{A}e^{(\theta-2\alpha)x^{+}}\partial_{-}f\ \delta_{AB}]
\cr
&e^{B}\wedge e^{-}
+(\theta-\alpha_{A}) e^{(\theta-2\alpha-\alpha_{A})x^{+}}\partial_{A}f e^{+}\wedge e^{-}
+\alpha_{A}e^{(\theta-2\alpha-\alpha_{B})x^{+}}\partial_{B}fe^{A}\wedge e^{B}\cr
&+\alpha_{A}(\alpha_{A}+\theta-2\alpha)e^{(\theta-4\alpha)x^{+}}f e^{A}\wedge e^{+},\cr
{\Omega^{A}} _{+} &=-\alpha_{A}(\alpha_{A}-2\alpha)\ e^{-4\alpha x^{+}}\ f\ e^{+}\wedge e^{-}
 +\alpha^{A}(\theta+\alpha^{A}-2\alpha)e^{(\theta-4\alpha)x^{+}}\  f\ e^{A}\wedge e^{-},\cr
{\Omega^{A}} _{B} &=\alpha_{A}\
e^{(\theta+\alpha_{A}-2\alpha)x^{+}}\ e^{A}\wedge e^{-}-\alpha_{B}
e^{(\theta+\alpha_{B}-2\alpha)x^{+}} e^{B}\wedge  e^{-}
-2\alpha_{A}\alpha_{B}
  e^{(\theta-4\alpha)x^{+}}\ e^{A}\wedge e^B.}}
  In the Lorentz frame, the non-vanishing components of the
Ricci tensor are \eqn\ricci{\eqalign{
         R_{++} &= -\sum_{i} \alpha_{A}(\alpha_{A}-2\alpha) e^{-4\alpha x^{+}},\cr
R_{--} &= -\sum_{A} \alpha_{A}(\alpha_{A}-2\alpha) e^{(2\theta-4\alpha) x^{+}}\ f^2\cr
         & +\sum_{A} e^{(\theta-2\alpha_{A})x^{+}} \partial_A\partial_A f
          +\sum_{A} \alpha_{A}e^{(\theta-2\alpha_{A})x^{+}}\ \partial_{-}f,\cr
         R_{+-}&=[\theta(\theta-2\alpha)+\sum_{A} \a_A(\c+\alpha_{A}-2\alpha)]\ e^{(\theta-4\alpha)x^{+}} f ,\cr
         R_{A+}&=0 ,\cr
          R_{A-} &=(\theta-2\alpha_{A}+\sum_{B} \alpha_{B})\ e^{(\theta-2\alpha-\alpha_{A})x^{+}} \partial_{A}f ,\cr
         R_{AB} &=-2\alpha_{A}(\theta+\sum_{C} \alpha_{C}-2\alpha)
           e^{(\theta-4\alpha)x^{+}}\ f\  \delta_{AB}.}}
Because the transformation from the orthonarmal basis to the coordinate
basis does not mix the spatial spatial components, we see from \ricci\ that
the condition $T_{AB}=0$ implies
\eqn\alone{
\theta-2\alpha+\sum_{A}\alpha_{A}=0.}
This condition also implies that $R_{A-}=0$ if $\a_A=\a$. This is the case for the
background \mmtm\ when $A=i$. If $\a_A\ne \a$, we will have to postulate $\p_A f=0$,
namely we need to smear the source in these directions. If we do not, a non-vanishing
component $T_{A-}$ is required, this is just fine, since $R_{A-}$ thus $T_{A-}$ is
a total derivative, the component of momentum
\eqn\momc{p_A\sim \int T_{A-}\sqrt{g_\bot}dx^-d^9x}
automatically vanishes. However, we will assume that as long as $\a_A\ne \a$,
the source is smeared in direction $x^A$. Finally,
Taking \alone\  as well as \eomf\ into account, and changing back to the coordinate basis, we
find the only non-vanishing component of Riccci tensor  $R_{--}$
\eqn\ano{
R_{--} =\sum_{A} e^{(\theta-2\alpha_{A})x^{+}} \partial_A\partial_A f
+\sum_{A} \alpha_{A}e^{(\theta-2\alpha_{A})x^{+}}\ \partial_{-}f.}

From now on we focus on the two cases \mtmet\ and \mmtm\ only. In
the case of \mtmet, $f$ is independent of $x^-$ and $Y$, and
\eqn\caone{\theta=-4Q/3.} The remaining Einstein equation is
\eqn\eins{ 8e^{-2Qx^+}\Delta f={\kappa_{11}}^2\sum_{a} p_{-a}
e^{-2Q x^+}\delta(\vec{x}_{\bot} -\vec{x}_{\bot a}),} where
$\Delta$ is the Laplace operator in the 8 transverse dimensions.
Finally we get the solution \eqn\solu{
f=-\sum_{a}{\kappa_{11}^2p_{-a}\over
48s_{7}|\vec{x}_{\bot}-\vec{x}_{\bot a}|^6},} where $s_7$ denotes
the volume of the sphere $S^7$.

The surface defined by $x^+=$ constant becomes spacelike now. So
we will get a spacelike singularity. Quantitatively, when $g_{--}$
and $g_{+-}$ become the same order, the geometry is severely
changed. Consider only one source, there is a radius:
\eqn\radius{r_{0}=e^{-\alpha
x^+}({48s^7\over\kappa_{11}p_{-}})^{1\over 6}} The condition $r\gg
r_{0}$ defines the asymptotical region. When $x^+\rightarrow \infty,
r_{0}\rightarrow \infty$. It seems dangerous. But there is some
subtlety here. First, When $x^+$ is finite, $r_{0}$ is also finite,
and we do not need to worry about the change as long as we are far
away from the source. As $x^+$ grows, $r_{0}$ vanishes, and so the
future is changed little. Second, if we want to detect the change of
background, we must introduce another observer. Here we put another
graviton into the spacetime. The change of background is reflected
by the change of action of the second graviton. As we will see
later, this picture breaks down as $x^+\rightarrow -\infty$. So even
the nature of the singularity is changed, we cannot detect the
change on the classical level. To further understand it, perhaps we
should go to matrix theory. It is important to ask whether matrix
theory is right here. In matrix theory, two gravitons correspond to
block diagonal parts of a matrix. They interact via the off diagonal
part. Spacetime emerges as we integrate out the off diagonal part.
When $x^+\rightarrow -\infty$, we reach the singularity in sugra,
and a week coupling super Yang-Mills theory in matrix.  As argued by
\csv, simply integrating out is not allowed. We must deal with the
whole non-Abelian degrees of freedom. To us, there is no sign of
breakdown of matrix theory. More rigorous matrix calculation is
needed to say anything more.

In a flat background as well as in a anti-de Sitter background,
stability of a perturbation amounts to requiring that the space-time
remains asymptotically flat or anti-de Sitter, since definition of
S-matrix or boundary correlators demands this. In a time-dependent
background with a big-bang singularity, intuitively, the space-time structure
is required to remain the same where there is no singularity as time
approaches infinity, since we may invoke perturbative string picture
there. However, at the point where there is singularity, the definition
of observables is not clear yet, so we do not know how to define stability.
Nevertheless, the kind of divergence found in \hp\ is not allowed, happily
that does not occur here.

Unlike the case discussed in \alor, the change of the geometry is
localized here. In the region far away from the sources, space-time
is almost the same as the original background. In this sense, the
background is stable. In the case of \mmtm, the calculation is
similar. Provided that we compactify $x^-$ and $x^a$ and smear the
source in these directions, all components of Ricci tensor vanish
except $R_{--}$. The solution of $f$ is also similar, $r^6$ in the
denominator in \solu\ is to be replaced by $r^{7-d}$. If we take the
radius of $x^-$ to infinity, the stress tensor should also be
localized in this direction, the source will contain a factor
$\delta(x^-)$, so does the component $g_{--}$ of the metric. $\delta
(x^-)$ is also localized, it will not trigger instability.

What we have obtained is noting other than the smeared shock wave representing the
classical background of multi D0-branes. Eq.\solu\ clearly indicates that there is
no net interaction among D0-branes.  We now discuss whether this background still
preserves supersymmetry.

\noindent $\bullet$ Supersymmetry

In 11 dimensions, the only SUSY transformation of interest is that for the gravitino
$\delta\psi_{\mu}=D_{\mu}\epsilon$. The components of spin connection of interest are
$\omega_{-}$ and $\omega_{i}$:
\eqn\spinc{\eqalign{ \omega_{-} &=2\theta e^{(\theta -2\alpha)x^{+}} f
  \Gamma^{+ -}+2\sum_{i}e^{(\theta-\alpha_{A})x_{+}} \partial_{A}f\ \Gamma^{A-}, \cr
  \omega_A&=2\a_A e^{(\a_A-2\a)x^{+}}\ \Gamma^{A+} -2\a_A e^{(\c +\a_A-2\a)x^{+}}\ f\ \Gamma^{A-}.}}
Then the condition $D_+\epsilon=0$ says that $\epsilon $ is independent of $x_{+}$.
For the condition
\eqn\dcon{D_-\epsilon=(\partial_{-}+{1\over 4} \omega_{-}) \epsilon=0}
multiplied by $\Gamma^{-}$, it becomes
\eqn\becom{
(\partial_{-}+\half\theta e^{(\theta-2\a)x^{+}}\ f) \Gamma^{-}\epsilon=0}
The second term is dependent of $x^{+}$, while the first term not.
Thus they both vanish, implying that $\epsilon$ is also independent of $x^{-}$,
and that $\epsilon$ is subjected to the constraint $\Gamma_{-}\epsilon=0$.
Similarly, multiply $D_A\epsilon=0$ by $\Gamma_{-}$,
we end up with  $\Gamma_{A}\epsilon=0$ for all transverse directions, $\epsilon$
has to vanish.
Hence we have shown that the existence of sources breaks all supersymmetry.

\noindent $\bullet$ Interaction

We now switch to the convention in which $x^{-}=x-t$, hence
the momentum $p_{-}$ is always positive, and $f$ is always
negative. Consider the case that there is only one source. The
solution looks much the same as the Aichelburg-Sexl metric. Hence
we follow the approach of \becker\ to discuss the probe action of
graviton of momentum $N_{2}/R$ in the background of graviton of
momentum $N_{1}/R$. When $N_{1}$ is large, the first graviton is
treated as a classical source and produces the metric we have got.
Then take the other particle as a probe particle. To get the action
of the probe graviton in this background, the trick is to begin
with the action for a massive scalar in eleven dimensions
\eqn\prob{\eqalign{S&=-m\int
d\tau(-G_{\mu\nu}\dot{x}^\mu\dot{x}^\nu)^\half\cr &=-m\int
d\tau(-2e^{2\a x^+}\dot{x}^--\sum_Ae^{2\a_A x^+}(\dot{x}^A)^2
+2e^{\c x^+}f \dot{x}^- \dot{x}^-)^\half.}} A dot denotes
$\partial_{\tau}$. Carry out a Legende transformation on $x^-$:
\eqn\legen{ p_{-}=m(e^{2\a x^+}-2e^{\c x^+}f\dot{x}^-)(2e^{2\a
x^+}\dot{x}^--e^{2\a_A x^+} v_A^2+2e^{\c
x^+}f\dot{x}^-\dot{x}^-)^{-\half}.} In our case $p_{-}$ is to be
fixed. Taking the limit $ m\rightarrow 0$ implies that
$G_{\mu\nu}\dot{x}^\mu\dot{x}^\nu \rightarrow 0$. Then \eqn\mvel{
\dot{x}^-={e^{2\a x^+}-\sqrt{e^{4\a
x^+}+2\sum_Ae^{(\c+2\a_A)x^+}(\dot{x}^A)^2}\over 2f e^{\c x^+}}.}
The appropriate Lagrangian is the Routhian, \eqn\rout{ {\cal
L}'(p_{-})=-{\cal R}(p_{-})={\cal L}-p_{-}\dot{x}^-(p_{-}).}

When $m\rightarrow 0$ with fixed $p_{-}$, \eqn\rout{\eqalign{{\cal
L}'(p_{-})&\rightarrow -p_{-}\dot{x}^-\cr
&=p_{-}\{\half\sum_{A}e^{2(\a_A-\a)x^+}(\dot{x}^A)^2-{1\over 4}
e^{-6\a+\c}[\sum_{A}e^{2\a_Ax^+}(\dot{x}^A)^2]^2 f\cr
&+O(f^2\dot{x}_{\bot}^4)\} .}} The interaction terms all involve
relative velocity $v$. If the source and the probe stay relatively
static in the transverse space, there is no net interaction. This
has to be the case if matrix model is correct, since two static
D0-branes form a solution in matrix model, and induces no
interaction terms in the matrix model action.

When the parameters take the values in \mone, we get
\eqn\nrout{\eqalign{ {\cal L}'(p_{-})&\rightarrow
p_{-}\{\half[\sum_{i}^8(\dot{x}^i)^2+e^{-2Qx^+}(\dot{x}^9)^2]\cr
&-{1\over 4}[\sum_{i}^8e^{-Qx^+}(\dot{x}^i)^2
+e^{-3Qx^+}(\dot{x}^9)^2]^2 f\cr &+O(f^2\dot{x}_{\bot}^4)\}.}} We
see that as $x^+\rightarrow -\infty$, those ``higher order" terms
become larger and larger, it is a bad idea to do expansion in
terms of the transverse velocity.

The double expansion in $v$ and $r^{-1}$ corresponds to
perturbative expansion in the matrix model, the breaking-down of this
double expansion implies that even when one is treating
interaction between two D0-branes, one should no do perturbative
calculation in matrix model. It is rather curious that this
breaking-down happens, since the matrix model or the matrix string
theory in question becomes weakly coupled at early times. Our
interpretation of this breaking-down is the following. Although
the theory becomes weakly-coupled, the picture of two separated
D0-branes becomes worse and worse at early times, since the
off-diagonal modes become lighter and lighter, this can be seen
for instance from the interaction term
\eqn\intt{Rg_s^{-2}\tr[X^i,X^j]^2=Re^{2Q t}\tr [X^i,X^j]^2,} where
$t=x^+$. The mass of the off-diagonal modes is proportional to $e^{Qt}$.
The breaking-down of velocity expansion is simply due to
appearance of new light modes. Thus, although the D0-brane
``moduli" (position) description breaks down, the matrix model
does not, one simply has to deal with all non-abelian degrees of
freedom. This is how matrix model resolves the singularity.

Acknowledgments.

This research was supported by a grant of CNSF, we thank Q. G. Huang and
Hong Lu for helpful
discussions. WS would like to thank Ke Ke for teaching her how to use Maple
to calculate the curvature tensor.

\listrefs
\end